\documentclass[twocolumn,aps,prl,showpacs,superscriptaddress]{revtex4-1}
\usepackage[latin9]{inputenc}
\usepackage{amsmath}
\usepackage[dvipdfmx]{graphicx}

\usepackage{lineno}

\makeatletter

\usepackage{verbatim}
\usepackage{color}
\usepackage{subfigure}

\makeatother

\begin{document}

\title{Observation of nonlocal Josephson effect on double InAs nanowires}

\author{Sadashige Matsuo}
\email{sadashige.matsuo@riken.jp}
\affiliation{Center for Emergent Matter Science, RIKEN, 2-1 Hirosawa, Wako-shi, Saitama 351-0198, Japan}
\affiliation{JST, PRESTO, 4-1-8 Honcho, Kawaguchi, Saitama 332-0012, Japan}

\author{Joon Sue Lee}
\affiliation{California NanoSystems Institute, University of California, Santa
Barbara, CA 93106, USA}
\affiliation{Department of Physics and Astronomy, University of Tennessee, Knoxville, Tennessee 37996, USA}

\author{Chien-Yuan Chang}
\affiliation{Center for Emergent Matter Science, RIKEN, 2-1 Hirosawa, Wako-shi, Saitama 351-0198, Japan}

\author{Yosuke Sato}
\affiliation{Center for Emergent Matter Science, RIKEN, 2-1 Hirosawa, Wako-shi, Saitama 351-0198, Japan}
\affiliation{Department of Applied Physics, University of Tokyo, 7-3-1 Hongo, Bunkyo-ku, Tokyo 113-8656, Japan}

\author{Kento Ueda}
\affiliation{Department of Applied Physics, University of Tokyo, 7-3-1 Hongo, Bunkyo-ku, Tokyo 113-8656, Japan}

\author{Christopher J. Palmstr{\o}m}
\affiliation{California NanoSystems Institute, University of California, Santa
Barbara, CA 93106, USA}
\affiliation{Electrical and Computer Engineering, University of California, Santa 
Barbara, CA 93106 USA}
\affiliation{Materials Department, University of California, Santa
Barbara, CA 93106, USA}

\author{Seigo Tarucha}
\email{tarucha@riken.jp}
\affiliation{Center for Emergent Matter Science, RIKEN, 2-1 Hirosawa, Wako-shi, Saitama 351-0198, Japan}
\affiliation{Department of Physics, Tokyo University of Science, 1-3 Kagurazaka, Shinjuku-ku, Tokyo 162-8601, Japan}

\begin{abstract}
Short-range coherent coupling of two Josephson junctions (JJs) are predicted to generate a supercurrent in one JJ nonlocally modulated by the phase difference in the other.
We report on observation of the nonlocal Josephson effect on double InAs nanowires as experimental evidence of the coherent coupling.
We measured one JJ sharing one superconducting electrode with the other JJ and observed switching current oscillation as a control of the nonlocal phase difference.
Our result is an important step toward engineering of novel superconducting phenomena with the short-range coherent coupling.
\end{abstract}
\maketitle 
The development of Josephson junction (JJ) physics~\cite{Josephson1962} is of significance for exploiting novel coherent macroscopic quantum phenomena and superconducting (SC) device applications in various quantum information technologies~\cite{Krantz2019}. 
In particular, coherent coupling between two JJs is the key for SC circuit designs to engineer qubit-qubit couplings and realise a novel SC phase in multiple JJ arrays. 
Recently, the Andreev molecular state (AMS), a new concept for producing short-range coherent coupling between two JJs on nanowires has been introduced~\cite{Pilletnl2019,Kornichprr2019, Kornichprb2020}. 
In the two adjacent JJs, upper (JJU) and lower (JJL), on the nanowires sharing one SC electrode, as described in Fig. 1(a), the respective Andreev bound states (ABSs)~\cite{Furusaki1991, beenakkerprl1991} are hybridised because of the penetration of the ABS wave functions into the shared SC electrode, forming an AMS as the bonding and anti-bonding states.
ABSs have been utilised for the Andreev qubit~\cite{Chtchelkatchev2003, Zazunov2003, Janvier2015}, and AMS physics holds the possibility of producing short-range coherent coupling between qubits. In such a device, the ABS energy and supercurrent in JJU depend on not only $\delta _U$, the phase difference on JJU, but also on the ABS energy and the phase difference, $\delta _L$, of JJL. Consequently, the supercurrent in JJU can be controlled nonlocally by manipulation of JJL, which is referred to as the nonlocal Josephson effect~\cite{Pilletnl2019}. The AMS has been observed in a double dot coupled with SCs~\cite{Kuertoessy2021}; however, experimental studies of superconducting transport related to the AMS, such as the nonlocal Josephson effect, have not been reported. 

When two nanowires are coherently coupled through an SC electrode, the coupling forms a nonlocal SC correlation on the double nanowire. Such nonlocal SC correlation is an essential ingredient for engineering time-reversal invariant topological SCs and can be applied in topological quantum computing with Majorana fermions or parafermions in SC-semiconductor hybrid systems~\cite{jelenaprb2014, Gaidamauskas2014, Haim2014}. The nonlocal SC correlation on double nanowires has been addressed in electrically tunable devices~\cite{Baba2018, Ueda2019, Kuertoessy2021}. 
Experimental study of the nonlocal Josephson effect is an important step to address the phase control of the nonlocal SC correlation; it paves the way for realising time-reversal invariant topological SC devices.

\begin{figure}[t]
\includegraphics[width=0.95\linewidth]{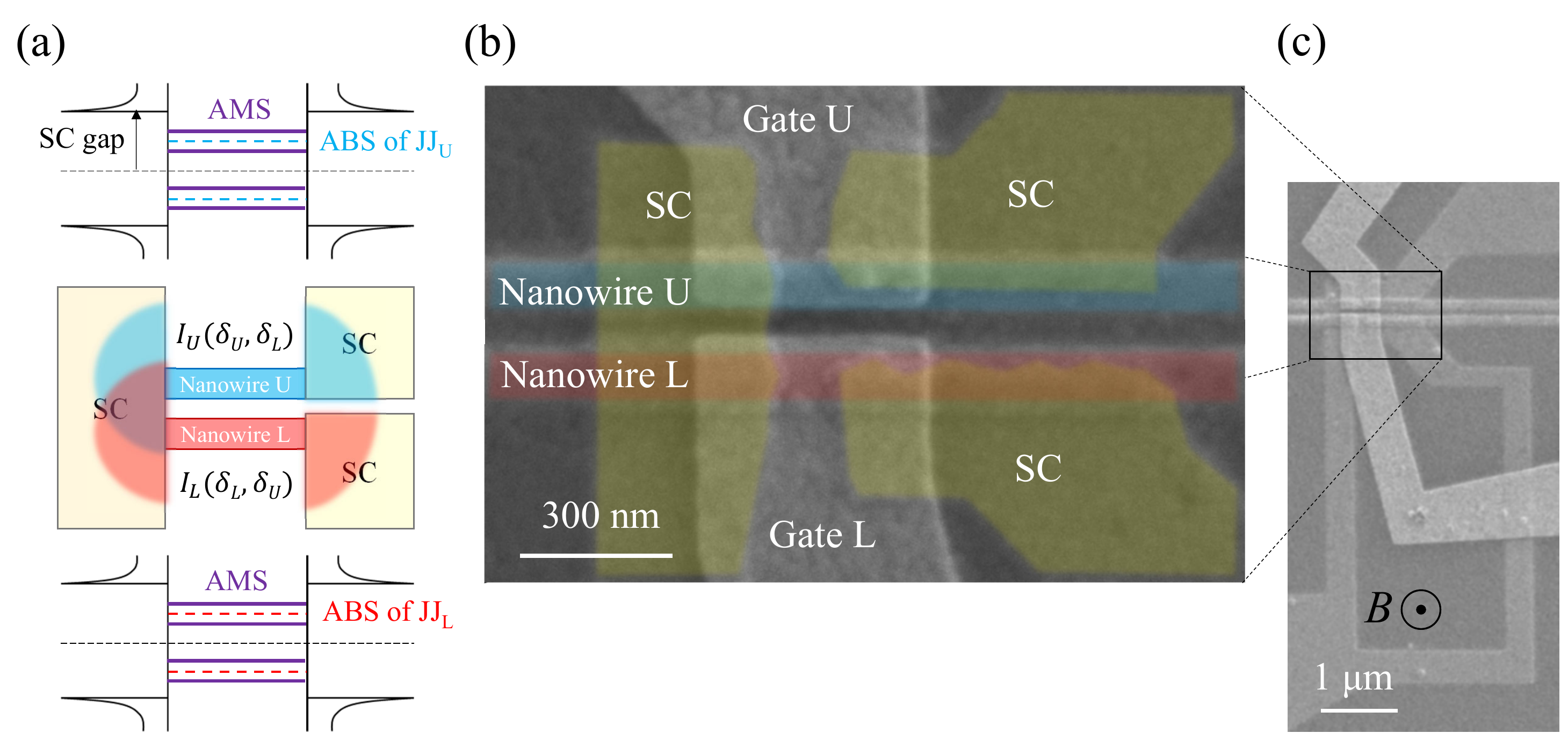}
\caption{{\bf Concept of this study and device structure.}
 (a) AMS formation is shown. The ABSs of JJU and JJL penetrate the SC electrodes, as illustrated in the middle panel. When the separation of the two JJs is short, the ABSs in the respective nanowires are hybridised to form an AMS. (b) SEM image of the 3-terminal JJ on a double nanowire. The two JJs share the left SC electrode, but on the right side, the two SC electrodes were placed on the respective  nanowires. (c) SEM image of sample B. The shared and lower SC electrodes in (b) are connected by a superconductor loop.}
\label{fig1} 
\end{figure}

Considering that multiple phase differences dominate the junction properties, the coupled JJs in Fig. 1(a) resemble multiterminal JJs~\cite{Freyn2011, Pfeffer2014, Feinberg2015, Strambini2016, Cohen2018, Draelos2019, Pankratova2020, Graziano2020}, providing a platform for topological physics~\cite{Heck2014, Yokoyama2015, Riwar2016, Xie2017}. In this sense, establishing the physics of the coupling between JJs will also help to develop multiterminal JJ physics, as studies on not only devices of a single normal conductor with multiple SC electrodes but also those of two or more coupled JJs on various independently controllable materials may be conducted. 

Here, we report experimental evidence of the nonlocal Josephson effect using gate-tunable InAs nanowires to demonstrate the coherent coupling between two JJs on a double nanowire. 
Our results provide a means to nonlocally control the D.C.  Josephson effect. 
Our strategy for this demonstration is to determine whether the SC correlation between any two SC electrodes exists and then measure the switching current in JJU dependent on $\delta _L$.

We prepared two coupled JJ devices, named samples A and B. We chose an InAs double-nanowire structure, created using the selective area growth (SAG) method as this method allows the preparation of two spatially separated parallel nanowires with high yield compared to using self-assembled nanowires~\cite{babaapl2017, Baba2018, Ueda2019}. Additionally, aluminium is epitaxially grown on InAs nanowires, which provides a highly transparent interface~\cite{Krogstrup2015}. The two SAG nanowires were spatially separated by 60 nm. JJU and JJL were formed on InAs nanowires U and L, respectively. The SC electrode on the left side is shared by the two JJs, but the two right SC electrodes contact nanowire U and nanowire L separately. Sample A was fabricated into a 3-terminal JJ device (left shared SC, right upper SC, and right lower SC electrodes) on nanowires U and L. A scanning electron microscope (SEM) image of the same structure is shown in Fig. 1(b). For sample B, JJL is embedded in an SC loop to change $\delta _L$ by a magnetic field ($B$) penetrating the SC loop. The two gate electrodes were used to electrically control JJU and JJL with gate voltages of $V_{\rm gU}$ and $V_{\rm gL}$, respectively. The SEM image of sample B is shown in Fig. 1(c). The junction of sample B has the same structure as that of sample A, but the shared and lower SC electrodes are connected to an SC loop. Sample A was used to confirm the SC correlation between any two SC electrodes and sample B to demonstrate the nonlocal control of the D.C. Josephson effect.

First, we measured sample A at 10 mK to study the correlation between the SC electrodes. For this purpose, we simultaneously measured $V_{\rm U}$ and $V_{\rm L}$, the voltages of the upper and lower SC electrodes with the shared SC grounded by sweeping the bias currents $I_{\rm U}$ and $I_{\rm L}$ through JJU and JJL, respectively (see Fig. 2(a)). Figs. 2(b) and (c) show the measured differential resistances $dV_{\rm U}/dI_{\rm U}$ and $dV_{\rm L}/dI_{\rm L}$ as functions of $I_{\rm U}$ and $I_{\rm L}$, respectively. In the square region of -30 nA$ < I_{\rm U} < 30$ nA and -60 nA$ < I_{\rm L} < 60 $ nA, $dV_{\rm U}/dI_{\rm U}$ and $dV_{\rm L}/dI_{\rm L}$ almost vanish in both figures, indicating that the supercurrent flows between the upper and the shared SC and between the lower and the shared SC. This means that the switching current is almost 30 nA for JJU and 60 nA for JJL. 

\begin{figure}[t]
\includegraphics[width=0.95\linewidth]{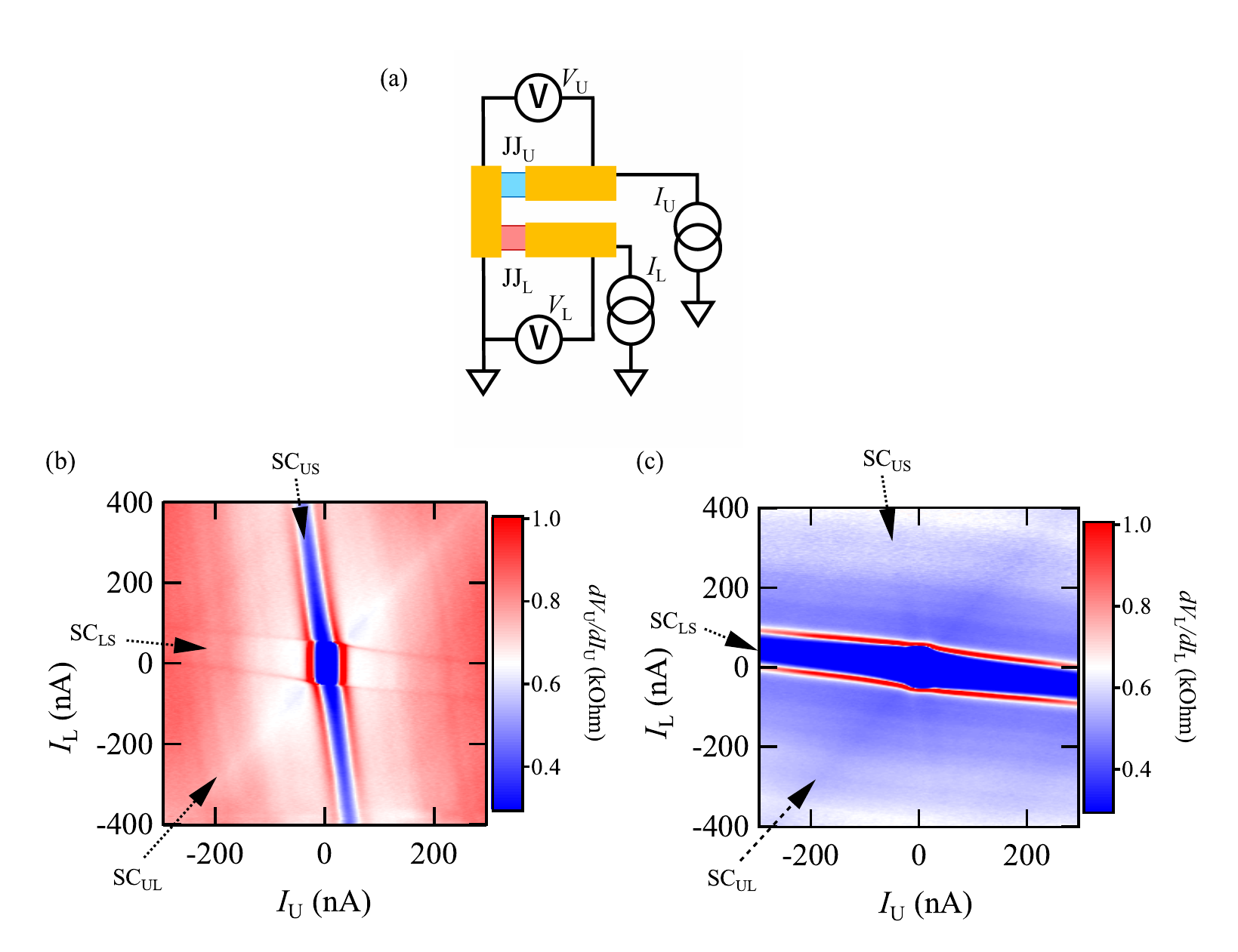}
\caption{{\bf 3 terminal measurement of the two coupled Josephson junctions on the double nanowire.}
(a) shows a schematic of the electrical circuit used for measurement of sample A. (b) and (c) show the differential resistances of JJU and JJL, respectively, as a function of $I_{\rm U}$ and $I_{\rm L}$, measured simultaneously. The supercurrent flows in both JJs in the central blue regions of -30 nA $< I_{\rm U} < 30$ nA and -60 nA $< I_{\rm L} < 60$ nA. In the regions arranged with ${\rm SC_{US}}$, ${\rm SC_{LS}}$, and ${\rm SC_{UL}}$ extending outward from the central blue supercurrent region, the local supercurrent remains between a pair of the two SC electrodes with the dissipative current in the other pairs.
}
\label{fig2} 
\end{figure}
From the supercurrent region, the three additional regions where the differential resistance is reduced are extended, as labelled with ${\rm SC_{US}, SC_{LS}, and SC_{UL}}$ in Figs. 2(b) and (c). These diagonal features have also been reported in multiterminal Josephson junctions~\cite{Draelos2019, Pankratova2020, Graziano2020}. In these regions, the supercurrent flowing between any two contacts remains within these contacts, whereas the dissipative current flows between the other contacts. 
For example, in ${\rm SC_{US}}$, there is considerable reduction, as seen in Fig. 2(b), but not in Fig. 2(c). In addition, ${\rm SC_{US}}$ is strongly dependent on $I_{\rm U}$ but is slightly sensitive to $I_{\rm L}$. This means that the supercurrent between the upper and shared SC electrodes remains in ${\rm SC_{US}}$ with the dissipative current at JJL because $|I_{\rm L}| > 60$ nA. For similar reasons, ${\rm SC_{LS}}$ can be assigned to the region where the supercurrent between the lower and shared SC electrodes remains.

As for ${\rm SC_{US}}$ and ${\rm SC_{LS}}$ regions, the figures indicate that the differential resistance in the remained supercurrent regions is finite. For example $dV_{\rm U}/dI_{\rm U}$ in ${\rm SC_{US}}$ is around 300 Ohm at $I_{\rm L} \sim \pm400$ nA (see supplemental material (SM)). This can be assigned to the Joule heating effect derived from the normal transport in JJL. The heating increases the electron temperature in JJU, resulting in the phase diffusion which produces the finite differential resistance even in the supercurrent region~\cite{Ambegaokar1969}. Additionally, ${\rm SC_{US}}$ and ${\rm SC_{LS}}$ have finite tilts on the $V_{\rm U}V_{\rm L}$ plane. These tilts can be derived from the cotunneling of the quasiparticles. For example, when a finite potential difference on JJU is present, the current in JJU flows from the upper SC to the shared SC. Additionally, due to the cotunneling through the shared SC, the finite current also flows to the lower SC. This cotunneling current generates the finite tilt of ${\rm SC_{LS}}$. 

We focus on the diagonal feature ${\rm SC_{UL}}$ in which both $dV_{\rm U}/dI_{\rm U}$ and $dV_{\rm L}/dI_{\rm L}$ are slightly reduced, as shown in Figs. 2(b) and (c). A supercurrent can exist in three pairs of upper, lower, and shared SC electrodes. The supercurrent in the upper and shared SC pair is assigned to ${\rm SC_{US}}$, and that in the lower and shared SC pair is assigned to ${\rm SC_{LS}}$. Therefore, ${\rm SC_{UL}}$ can be assigned to the supercurrent between the upper and lower SC electrodes. This result indicates that a nontrivial SC correlation exists between the upper and lower SC electrodes, although the shared SC electrode only intermediates two nanowires; no other material connects the two SC electrodes. 
This is a signature of the nonlocal SC correlation between the two Josephson junctions. However, the ${\rm SC_{UL}}$ signal is small and vague because the dissipative current and nonequilibrium quasiparticles coexist with the supercurrent. Therefore, a detailed discussion of the nonlocal SC correlation from these 3-terminal results is difficult.

\begin{figure}[t]
\includegraphics[width=0.95\linewidth]{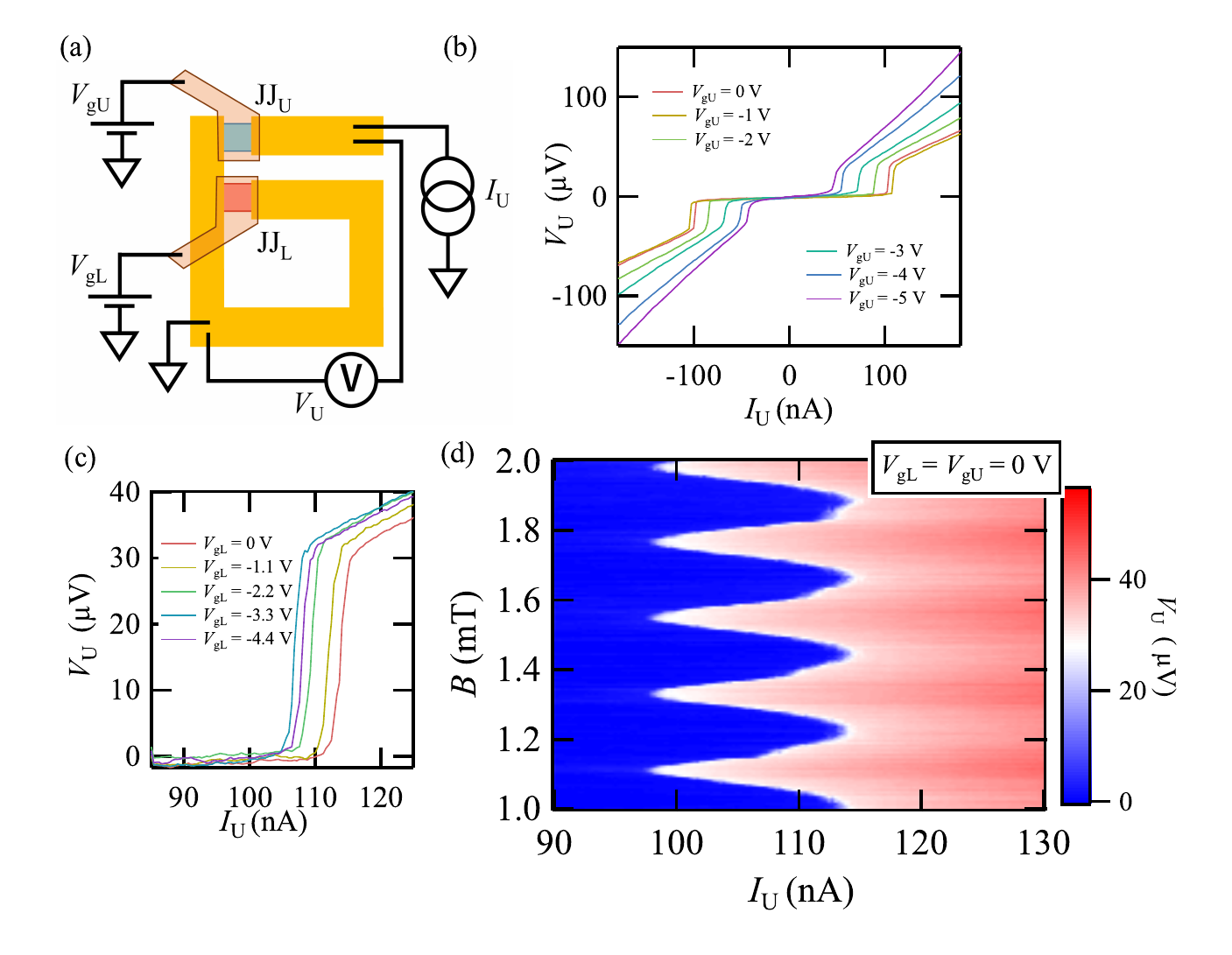}
\caption{{\bf Switching current oscillation indicating a coherent coupling between two Josephson junctions.}
(a) shows a schematic of the measurement circuit for sample B . (b) illustrates the $I_{\rm U}-V_{\rm U}$ curves obtained at several $V_{\rm gU}$ points, indicating that the supercurrent in JJU can be tuned by $V_{\rm gU}$ locally. 
(c) indicates $V_{\rm U}$ vs. $I_{\rm U}$ at $B=1.9$ mT and several $V_{\rm gL}$. The supercurrent in JJU is little controllable by $V_{\rm gL}$.
(d) shows $V_{\rm U}$ as a function of $I_{\rm U}$ and $B$. The boundary between the red and blue regions corresponds to the switching current in JJU as a function of $B$. The image clearly indicates the oscillation of the switching current in the JJU.
}
\label{fig3} 
\end{figure}
Then, we use another strategy to control the phase difference $\delta _L$ in sample B. We measured $V_{\rm U}$ by changing $I_{\rm U}$ and $B$ at several $V_{\rm gU}$ and $V_{\rm gL}$, the gate voltages on JJU and JJL respectively (see Fig. 3(a)). Fig. 3(b) represents $V_{\rm U}$ vs. $I_{\rm U}$ at $B=0$ mT and several $V_{\rm gU}$. At $I_{\rm U} = 105 $ nA for $V_{\rm gU}=0$ V, the supercurrent region is switched to the normal region. The switching current $I_{\rm sw}$ decreases as $V_{\rm gU}$ decreases, indicating that JJU can be controlled electrically. 
$V_{\rm gU}=-5$ V provides about half of $I_{\rm sw}$ at $V_{\rm gU}=0$ V. Figure 3(c) indicates $V_{\rm U}$ vs. $I_{\rm U}$ at $B=1.9$ mT and several $V_{\rm gL}$. $V_{\rm gL}=-4.4$ V provides about 0.9 of $I_{\rm sw}$ at $V_{\rm gL}=0$ V, meaning that $I_{\rm sw}$ of JJU depends weakly on $V_{\rm gL}$ in the present range. 

We measured $V_{\rm U}$ vs. $I_{\rm U}$ at various magnetic field strengths. The obtained values of $V_{\rm U}$ as a function of $I_{\rm U}$ and $B$ are shown in Fig. 3(d). As illustrated, the switching current around 100 nA, corresponding to the boundary between the blue and red regions, oscillates with $B$. The oscillation period is 0.22 mT, which reasonably agrees with the calculated value of 0.30 mT, derived from the superconducting loop area (6.88 ${\rm \mu m^2}$). The magnetic field changes only $\delta _L$ and the phase difference of JJL in the SC loop. Therefore, $B$ would not affect the $I_{\rm sw}$ of JJU if there were no nonlocal SC correlation between JJU and JJL. The obtained switching current oscillation was reproduced in different devices on a single or double nanowire (see SM). Furthermore, we confirmed that this oscillation disappeared as the distance between the two junctions increased, as expected from the theory that the distance should be shorter than the coherence length of SC to form the coupling.
Disappearance of the oscillation with the long distance implies that the single JJ on the single nanowire produces no oscillation of the switching current  (see SM). Consequently, we present the observation of the nonlocal Josephson effect derived from the coupling between JJU and JJL through the shared SC electrode. Additionally, our observation of JJU switching current in sample B proves coherence of the coupling because the nonlocal phase modulation is observed on the supercurrent which is one of the phase coherent phenomena. 

To clarify that the $I_{\rm sw}$ oscillation originates not only from the JJU, but also from the properties of JJL, we studied the gate voltage control of the nonlocal Josephson effect signal. We evaluated the peak-to-peak value and the average of $I_{\rm sw}$ vs. $B$ at the respective gate voltages. First, we indicate the $V_{\rm gU}$ dependence of $I_{\rm sw}$ vs. $B$ with $V_{\rm gL}=0$ V, indicating local gate control. Both the peak-to-peak value and the average of $I_{\rm sw}$ change with $V_{\rm gU}$, as shown in Fig. 4(a). Then, we move onto the $V_{\rm gL}$ dependence with $V_{\rm gU}=0$ V; this refers to nonlocal gate control. As shown in Fig. 4(b), the average of $I_{\rm sw}$ is almost constant, whereas the peak-to-peak value decreases as $V_{\rm gL}$ decreases (see SM). This indicates that the average is affected only by the local gate voltage; the oscillation is affected by both the local and nonlocal gate voltages. Therefore, the oscillation originates from the coherent coupling between the two JJs. 

\begin{figure}[t]
\includegraphics[width=0.95\linewidth]{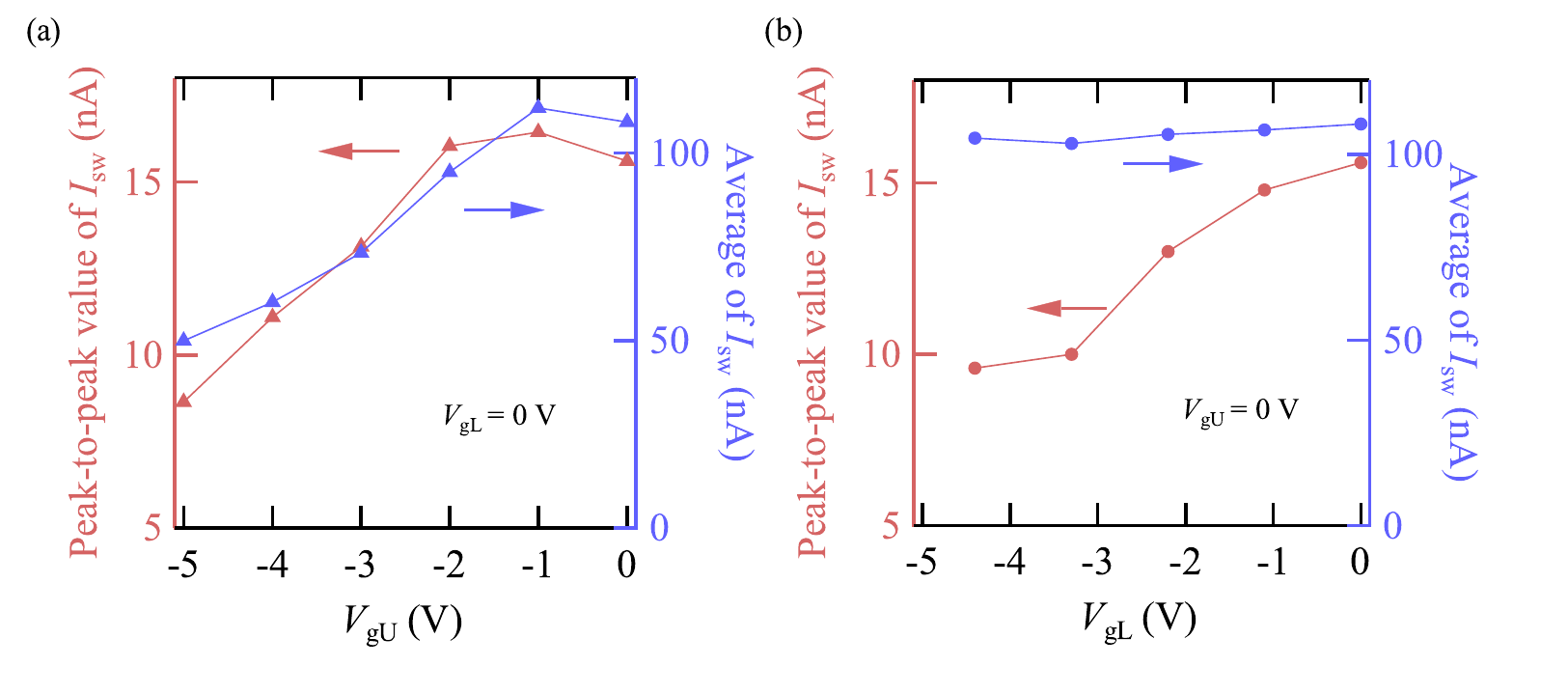}
\caption{{\bf Local and nonlocal gate control of the switching current oscillation in JJU.}
(a) and (b) show the peak-to-peak value in red and the average in blue of the switching current in JJU as a function of $V_{\rm gU}$ and $V_{\rm gL}$, respectively. In the local gate control case in (a), both the amplitude and average change with $V_{\rm gU}$. However, only the peak-to-peak value largely varies in the non-local gate control case in (b). This means that the switching current oscillation is derived from the hybridisation between JJU and JJL.
}
\label{fig4} 
\end{figure}
Finally, we discuss the microscopic origin of the observed coherent coupling between JJU and JJL. In the literature, the coupling in the ballistic junctions is dominantly formed by the double elastic cotunneling through the shared SC. The observed tilt of ${\rm SC_{LS}}$ in Fig. 2 provides a ratio of the cotunneling current to the current in JJU as around $0.14$. This value is similar to a ratio of the peak to peak value to the average of $I_{\rm sw}$ found in Fig. 4. Therefore, the similarity between the cotunneling ratio and the oscillation ratio seems to imply that the coherent coupling in our devices is dominantly originated from the cotunneling process as theoretically expected in the ballistic JJs.

Although our experiments clearly provide evidence of existence of the nonlocal Josephson effect, we only addressed the nonlocal control of the switching current.
In order to unveil the local and nonlocal transport on the coupled JJs, it is significant to evaluate the current phase relation, which gives the supercurrent as a function of local and nonlocal phase differences.
In addition, microscopic mechanism to produce the coherent coupling should be revealed experimentally.
Further studies on the coupled JJs are necessary.

In summary, we confirmed the nonlocal superconducting correlation between two JJs in a 3-terminal device and observed the nonlocal Josephson effect in a double JJ device with a superconducting loop. These demonstrated that the short-range coherent coupling between the two different Josephson junctions can be formed through a superconductor. 
The observed coherent coupling is not just for the semiconductor-superconductor junctions but expected in various kinds of Josephson junctions coupled through superconductors.
Our results pave the way to engineering short-range coherent coupling between superconducting qubits, such as the Andreev qubit~\cite{Chtchelkatchev2003, Zazunov2003, Janvier2015}, and provide a new building block for designing novel SC devices consisting of the two JJs on the different materials. 

\section*{Acknowledgments}
We thank P. Stano, C.-H. Hsu, Y. Tanaka, and S. Nakosai for fruitful discussions 
Funding: 
This work was partially supported by 
a JSPS Grant-in-Aid for Scientific Research (S) (Grant No. JP19H05610);
JST PRESTO (grant no. JPMJPR18L8);
Advanced Technology Institute Research Grants;
Ozawa-Yoshikawa Memorial Electronics Research Foundation; 
the US Department of Energy under Award No. DE-SC0019274.

\clearpage 

\section*{Supplementary Information}
\subsection*{Supplementary Note 1: Normal transport in sample B}
Before starting the condensation and circulation of the dilution fridge, we measured sample B  at 4 K. 
At this temperature, aluminium is not superconductive, and we can examine the normal transport properties of sample B. 
Figure S1 (a) shows the conductance $G$ in units of $2e^2/h$ as a function of $V_{\rm gL}$ and $V_{\rm gU}$. 
The line profiles are shown in Fig. S1 (b). 
The normal transport result has a small cross-capacitance effect between the upper and lower gate electrodes. 
This cross-capacitance effect can cause $V_{\rm gL}$ to change the carrier density of the nanowire U, which results in a small change in the average of $I_{\rm sw}$ as a function of $V_{\rm gL}$, as shown in Fig. 4 (b). 
In addition, we studied the resistance as a function of the magnetic field at 4 K, as shown in Fig. S1 (c). 
The result includes a small fluctuation but no clear periodic oscillation, like that seen in the switching current measurement shown in Fig. 3 (c). 
Then, the switching current oscillation obtained at the base temperature of our dilution fridge can be assigned to the hybridization of the ABSs in the two JJs, not to the normal state properties. 

\subsection*{Supplementary Note 2: Reproducibility in sample C}
We measured other selective area growth (SAG) double-nanowire devices with the same structure as sample B (two Josephson junctions on the double nanowire, one of which is embedded in a superconducting loop). 
We named this device sample C and repeated the measurements. 
Figure S2 represents $V_{\rm U}$ as a function of $I_{\rm U}$ and $B$ in sample C. 
An oscillating switching current similar to that in Fig. 3(c) was observed, as shown in Fig. S2. 
This result indicates the reproducibility of the result obtained for sample B.

\subsection*{Supplementary Note 3: Nonlocal Josephson effect on a single nanowire}
To study the nonlocal Josephson effect through the proximitized region, we fabricated a double JJ device on a SAG InAs single nanowire.
The scanning electron microscope (SEM) image is shown in Fig. S3. 
JJ1 is embedded in a superconducting loop, whereas JJ2 is out of the loop. 
The distance between the two JJs is 200 nm. 
We measured two samples, D and E, which have the same JJ structure, but with sample E having gate electrodes on the respective junctions. 
Figure S4 indicates $V_{\rm 2}$ of the voltage on JJ2 as a function of $I_{\rm 2}$, the bias current on JJ2, and $B$ of sample D. 
The result clearly shows the switching current oscillation, as seen in Fig. 3 in the main text. 
We observed a similar oscillation in sample E, as shown in Fig. S5 (a). 
We checked the gate voltage dependence of the switching current and evaluated the peak-to-peak value and the average of the oscillation as a function of $V_{\rm g1}$ and $V_{\rm g2}$, which represent the gate voltages on JJ1 and JJ2 as seen in Fig. S5 (b) and (c), respectively. 
The $I_{\rm sw}$ evolution with $V_{\rm g1}$ and $V_{\rm g2}$ has the same relation as that in Fig. 4(a) and (b).
The local gate voltage changes both the peak-to-peak value and the average, while the nonlocal gate voltage changes only the peak-to-peak value. 
 These results indicate that the nonlocal Josephson effect is generally obtained in the two coupled JJs on the semiconductor nanowires.
 
\subsection*{Supplementary Note 4: Dependence on the separation between two JJs on a single nanowire}
When the AMS is formed on the two JJs, the correlation is expected to vanish when the separation between two JJs is greater than the coherence length of the superconductivity between the two JJs. 
To measure whether the oscillations vanish in a device with a long separation between JJs, we fabricated four devices with the same structure as sample F, but with different separations of 0.2, 0.4, 0.6, and 1.0 ${\rm \mu m}$, on a single SAG nanowire.
The SEM image is shown in Fig. S6. 
The observed $I_{\rm sw}$ values of the coupled JJ devices with different separations $L$ are shown in Fig. S7(a). 
We note that the quality of the SAG nanowire is inhomogeneous even in a single nanowire; therefore we cannot ensure that all JJs have the same properties. 
However, the results indicate that a longer distance between two JJs produces less or no oscillation of the switching current. 
We executed the additional measurement on the single nanowire device with 150 nm separation on the different wafer as shown in Fig. S7(b). As a result, we found bigger amplitude of the oscillation whose ratio to the Isw average is about 0.2. These results indicate that the longer separation device tends to generate less oscillation.

The oscillation vanishes in devices of length 0.6  and 1.0 ${\rm \mu m}$. 
In this study of the length dependence, we used the single nanowire samples. Therefore, the two JJs are coupled through the proximity region beneath the shared SC. The coherence length is modified as the formula of $\hbar v_{\rm F}/\pi \Delta$ where $v_{\rm F}$ and $\Delta$ indicate the Fermi velocity and the superconducting energy, respectively. Since we cannot pinch off the nanowires due to the interfacial problem, we are not sure how large the carrier density and the Fermi velocity are. However, around 600 nm in the similar InAs/Al systems was reported previously~\cite{Mayer2019, Mayer2020}. This length is comparable to the shortest distance between the two JJs in our devices which did not show the oscillation. Then this length dependence is qualitatively consistent with AMS physics that the nonlocal SC correlation disappears when two JJs are at a larger distance than the coherence length. 

We note that the selective area growth nanowires are not uniform and the Josephson junction quality highly depends on the nanowires and also the junction location on the single nanowire. In case the grain boundary or impurities happen to appear in the junction, the junction quality can easily become small. Therefore, we assume that the general trend of decreasing modulation amplitude with increasing distance between nanowires is consistent between the prediction and experiment but only qualitatively.

\subsection*{Supplementary Note 5: Sample growth}
The SAG InAs double nanowires were grown by molecular beam epitaxy (MBE)~\cite{Schmid2015, Krizek2018, Friedl2018, Aseev2018,Vaitiekenas2018, Lee2019}. 
Prior to the InAs growth, nanowire structures were patterned using standard electron-beam lithography on ${\rm SiO_x}$ (30 nm)/${\rm Al_2O_3}$ (3 nm) layers on an Fe-doped semi-insulating InP(001) substrate. 
The patterns were formed by dry-etching of the ${\rm SiO_x}$ masking layer and wet-etching of the ${\rm Al_2O_3}$ etch-stop layer. 
The patterned nanowires are along the [100] ridge direction. 
The processed InP surface was further cleaned by ultraviolet-ozone and diluted HCl~\cite{Lee2019}. 
In a III-V MBE chamber, the patterned InP substrate was desorbed at 520\textdegree C with a flux of ${\rm As_4}$ molecules. 80-nm-thick InAs nanowires were grown at a growth rate of 0.1 ML/second  under As-rich growth conditions. 
After InAs growth, the substrate was transferred to an interconnected MBE chamber and cooled to $\sim$80 K using liquid nitrogen on a cold-head manipulator for Al layer growth. 
To achieve a continuous Al layer between two adjacent InAs nanowires along the [100] ridge direction, a 10-nm-thick Al layer was first grown at +20\textdegree \ with respect to the [100] ridge direction, and another 10-nm-thick Al layer was grown at -20\textdegree \ with respect to the [100] ridge direction. 
Immediately after Al growth, the substrate was transferred to a loadlock within 5-6 min and exposed to ${\rm O_2}$ gas for oxidation of the Al surface to preserve a smooth and continuous Al layer prepared at a cryogenic temperature. 

\subsection*{Supplementary Note 6: Device fabrication}
Epitaxial aluminium is grown globally on the substrate, and the two parallel nanowires are intermediated by the Al film. In the fabrication process, we removed the unnecessary epitaxial aluminium by wet etching with a type-D aluminium etchant to form JJs and superconducting loops. Then, we grew a 20 nm-thick aluminium oxide layer through atomic layer deposition and fabricated a separate gate electrode on each nanowire with Ti 5 nm and Au 20 nm. Following the measurement, we checked that the upper and lower SC electrodes were disconnected by SEM. 
Due to the interface quality of InP substrate and InAs nanowire~\cite{Krizek2018}, we cannot pinch off the conductance by the top gate voltages.

\subsection*{Supplementary Note 7: Measurement}
For the measurement of sample A, we injected bias currents from the upper and lower SC electrodes, with the shared SC electrode grounded. When we measured $dV_{\rm U}/dI_{\rm U}$ ($dV_{\rm L}/dI_{\rm L}$), we added a small oscillating current of 1 nA $\Delta I_{\rm U} (\Delta I_{\rm L})$ with 23 Hz on $I_{\rm U} (I_{\rm L})$ and detected the oscillation component of the voltage $\Delta V_{\rm U} (\Delta V_{\rm L})$ using the lock-in technique. For the measurement of sample B, we only injected a DC bias current $I_{\rm U}$ and detected $V_{\rm U}$. All measurements were performed at 10 mK, the base temperature of our CryoConcept wet dilution fridge.

\subsection*{Supplementary Note 8 : IV curves}
In our device, the normal resistance is a few hundred Ohm, and the gate structure works as the parallel capacitance. In the RCSJ model~\cite{Stewart1968, McCumber1968} which has been utilized for the Josephson junction analysis, $\beta$ (McCumber parameter) term is an important parameter to determine if the junction is overdamped or underdamped. When the beta is highly smaller than 1, then the junction is overdamped, and hysteresis in the IV curve does not appear. The estimated $\beta$ in sample B is around 0.025 at $B=0$ mT which means the junction is highly overdamped. Therefore, we did not observe the clear hysteresis in our study. 

\subsection*{Supplementary Note 9 :Line profiles of Fig. 2}
We exhibit several line profiles of Figs. 2(b) and (c) on the main text. Figures S8(a) and (b) indicate the line profiles on Fig. 2(b) at $I_{\rm L}=\pm400, \pm200, 0$ nA and on Fig. 2(c) at $I_{\rm U}=\pm200, \pm100, 0$ nA. At $I_{\rm L}=0$ nA, $dV_{\rm U}/dI_{\rm U}$ is not perfectly 0 nA in Fig. S8(a). This is because of the phase diffusion~\cite{Ambegaokar1969}. As $I_{\rm L}$ becomes far from 0 nA, suppression of $dV_{\rm U}/dI_{\rm U}$ corresponding to ${\rm SC_{US}}$ becomes smaller. This is assigned to the Joule heating effect on JJL accelerating the phase diffusion in JJU. The same features can be seen in Fig. S8(b).

\clearpage
\begin{figure}[t]
\includegraphics[width=0.65\linewidth]{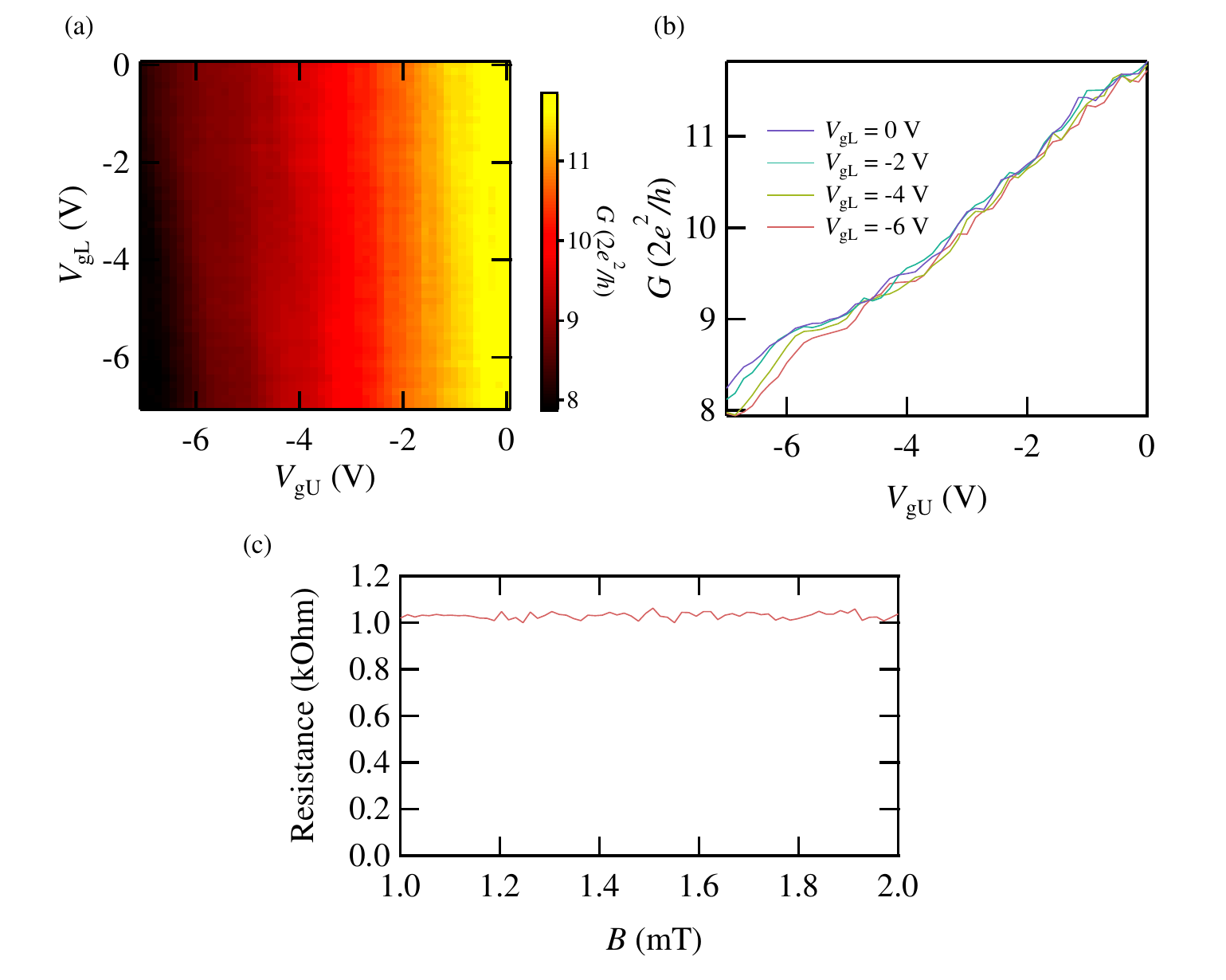}
\caption{
(a) shows the conductance as a function of $V_{\rm gL}$ and $V_{\rm gU}$ of sample B measured at 4 K. The line profiles at various VgLs appear in (b). The small measured $V_{\rm gL}$ dependence can be attributed to the cross-capacitance contribution. (c) Resistance as a function of $B$ at 4 K. 
There appear small fluctuations but no periodic oscillation.
}
\end{figure}

\begin{figure}[t]
\includegraphics[width=0.65\linewidth]{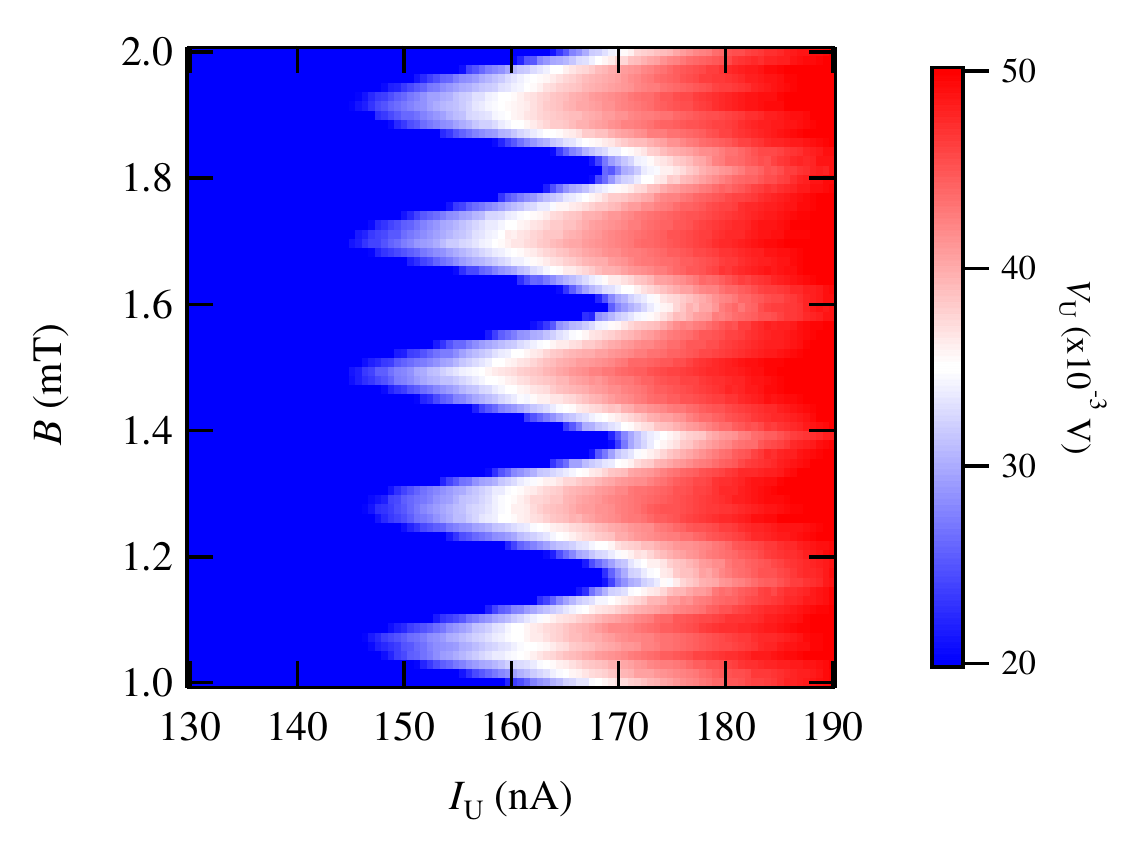}
\caption{
$V_{\rm U}$ as a function of $I_{\rm U}$ and $B$ obtained from sample C. A clear oscillation, as seen in Fig. 3 (c), was measured.
}
\end{figure}

\begin{figure}[t]
\includegraphics[width=0.45\linewidth]{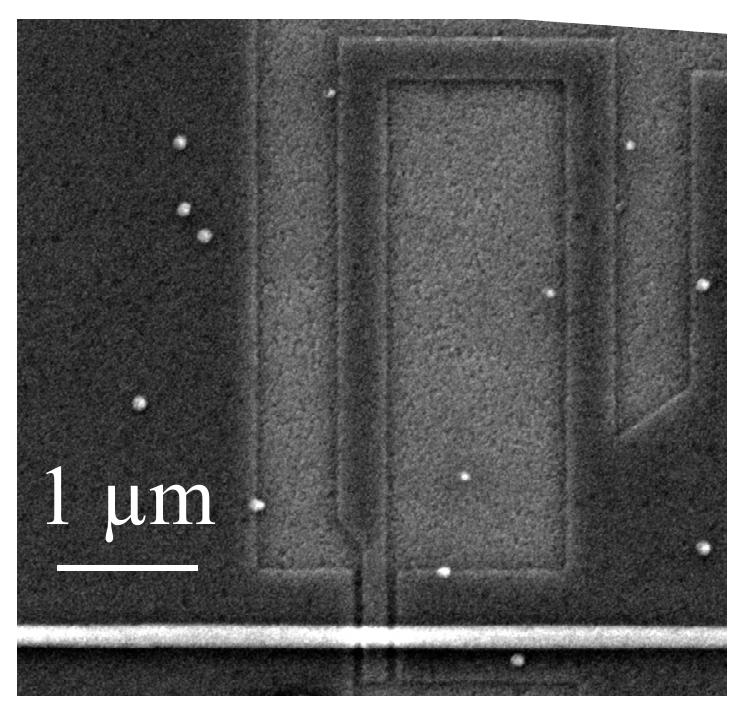}
\caption{
SEM image of the double Josephson junction on an SAG single nanowire. The two JJs were separated at  200 nm. The JJ length is 80 nm.
}
\end{figure}

\begin{figure}[t]
\includegraphics[width=0.75\linewidth]{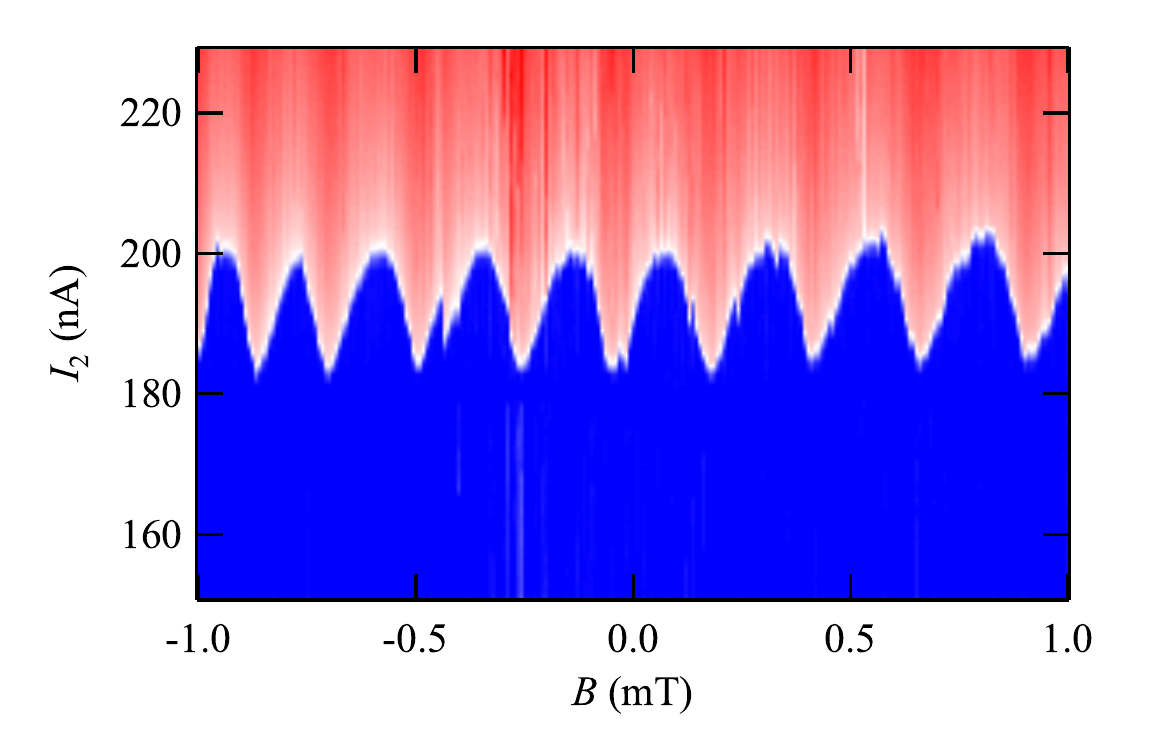}
\caption{
Switching current of the sample D oscillates as a function of $B$.
}
\end{figure}

\begin{figure}[t]
\includegraphics[width=0.75\linewidth]{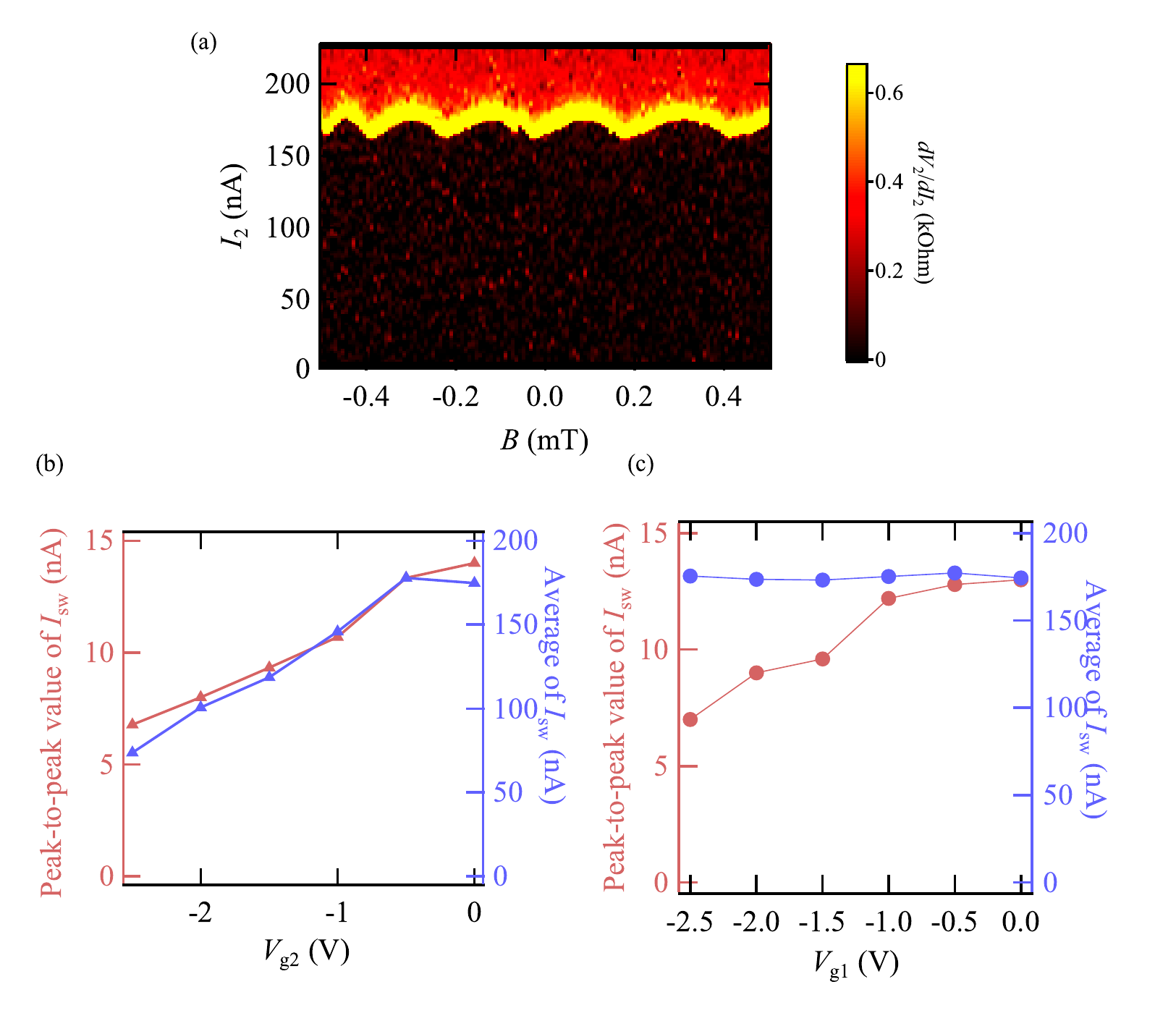}
\caption{
(a) indicates the differential resistance of sample E as a function of $B$ and $I_{\rm 2}$. Oscillation is observed as in Fig. 3 (c); the peak-to-peak value and the average hold the local and nonlocal gate voltage dependence similar to that in Figs. 4 (a) and (b). 
}
\end{figure}

\begin{figure}[t]
\includegraphics[width=0.65\linewidth]{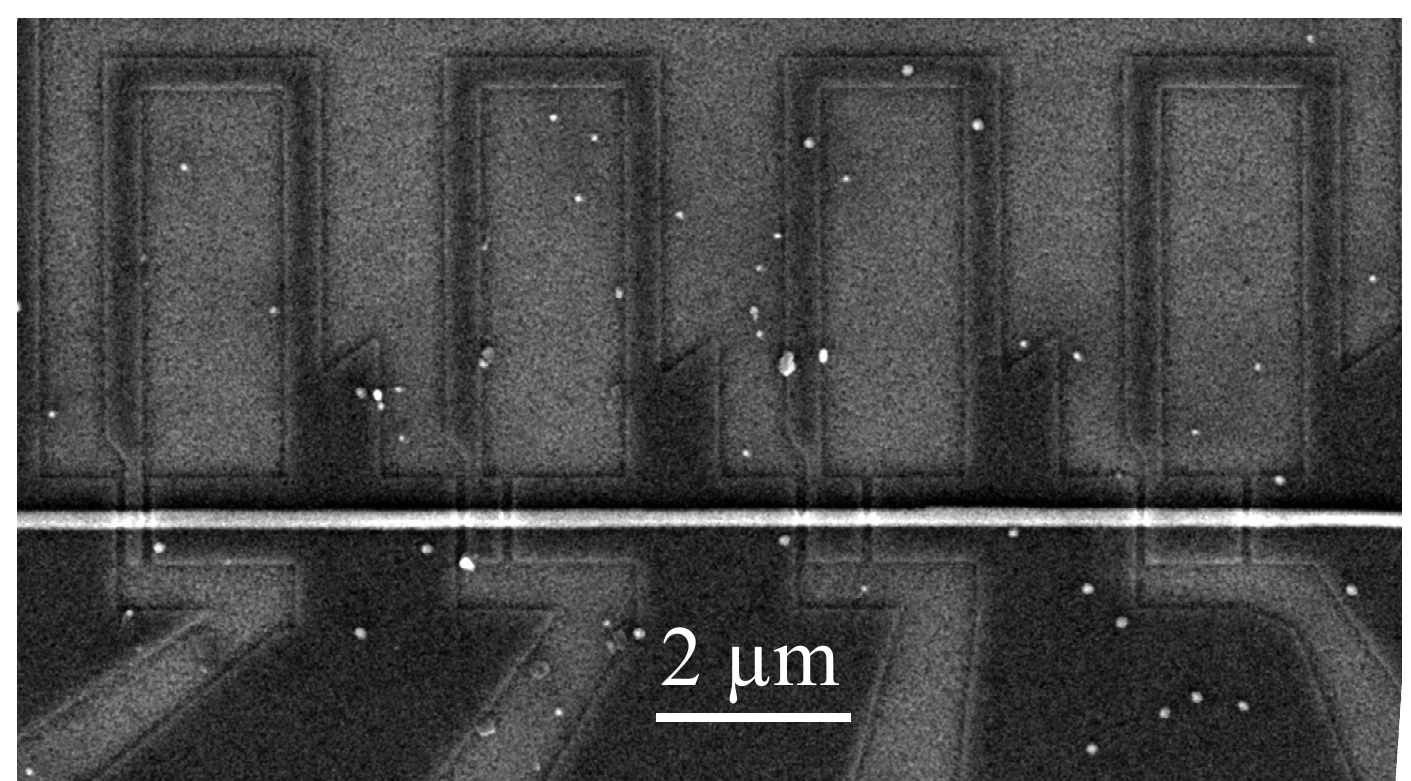}
\caption{
SEM image of double JJs on single nanowire with different separations. From left to right, the separations were 0.2, 0.4, 0.6, and 1.0 ${\rm \mu m}$.
}
\end{figure} 

\begin{figure}[t]
\includegraphics[width=0.85\linewidth]{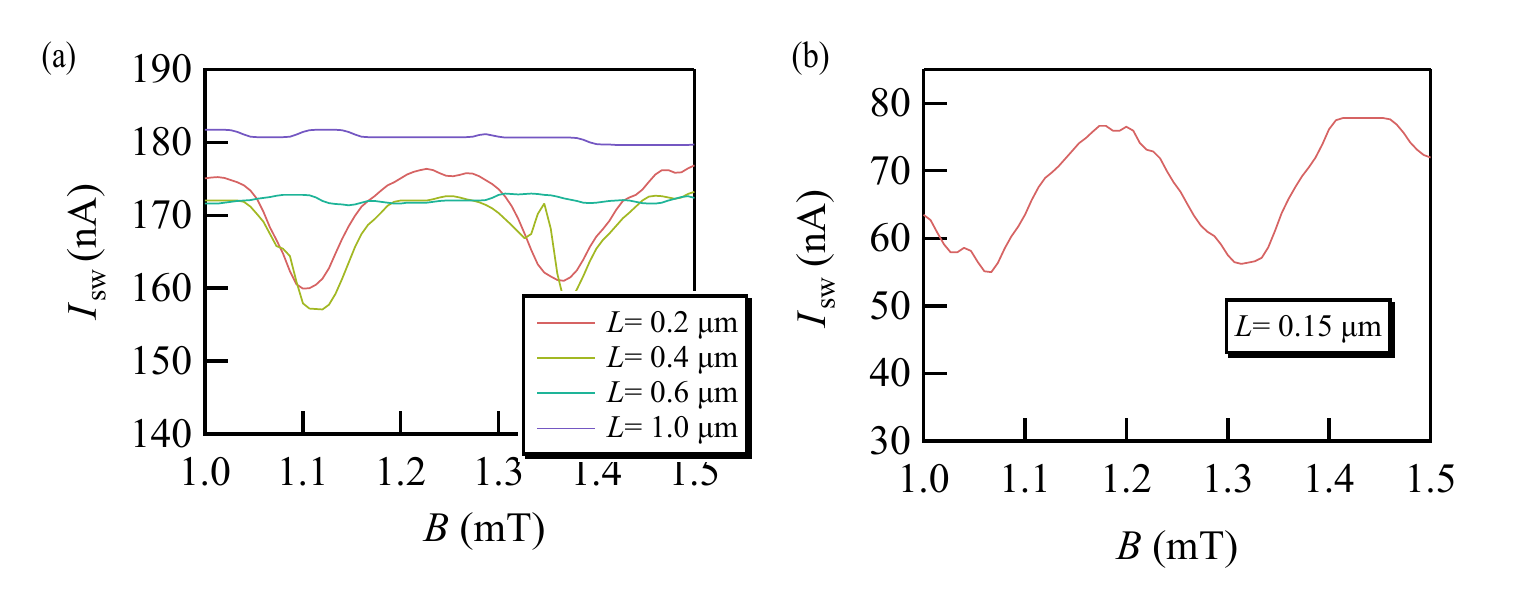}
\caption{
(a)$I_{\rm sw}$ as a function of $B$ of the coupled JJ devices with different separations in sample F is represented. Coupled JJ devices with separation longer than 0.6 ${\rm \mu m}$ have no oscillation.
(b)$I_{\rm sw}$ as a function of $B$ of the coupled JJ devices with 150 nm separation is represented.
}
\end{figure}

\begin{figure}[t]
\includegraphics[width=0.85\linewidth]{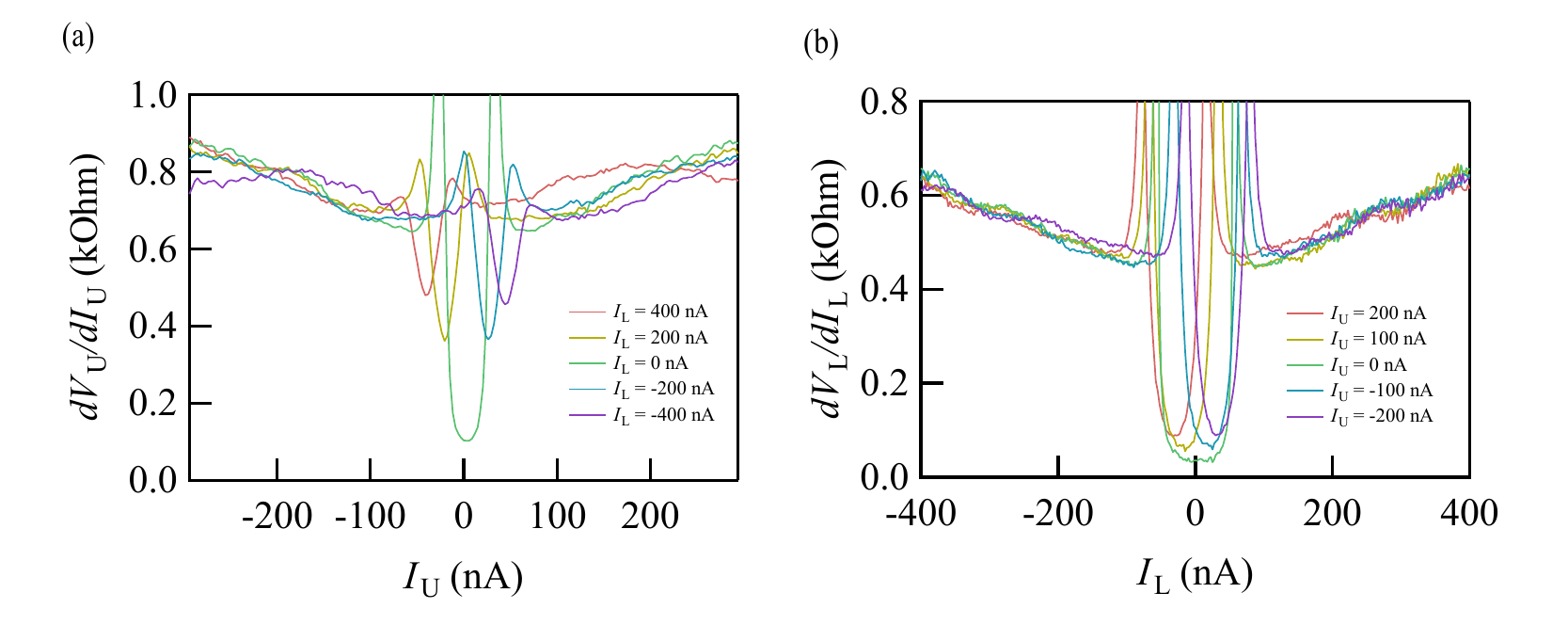}
\caption{
(a)The line profiles of Fig. 2(b) at $I_{\rm L}=\pm400, \pm200, 0$ nA.
(b)The line profiles of Fig. 2(c) at $I_{\rm U}=\pm200, \pm100, 0$ nA.
}
\end{figure}

\begin{figure}[t]
\includegraphics[width=0.85\linewidth]{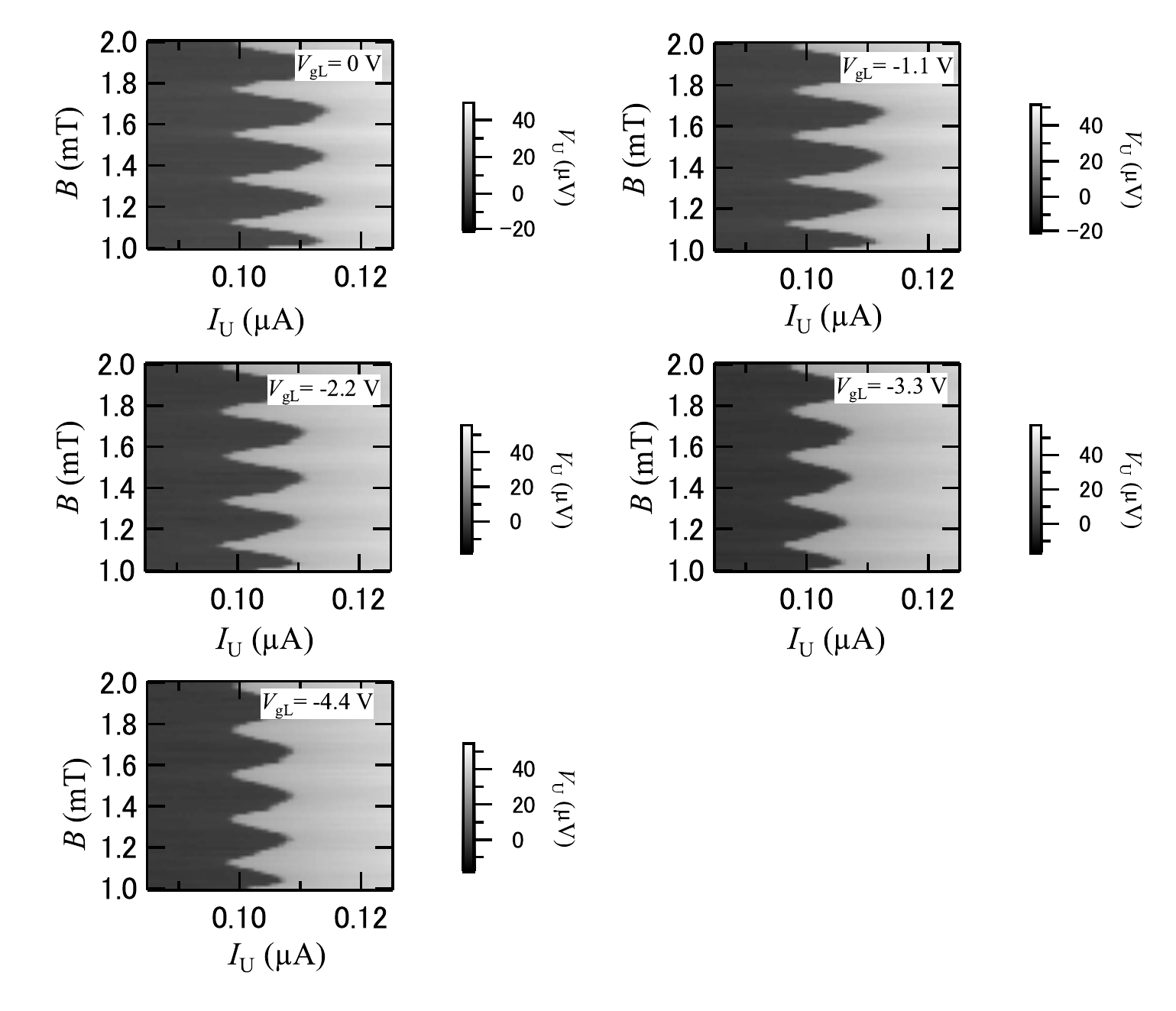}
\caption{
$V_{\rm U}$ results as a function of $B$ and $I_{\rm U}$ at several $V_{\rm gL}$ are shown. From these panels, we estimated the switching current and constructed Fig. 4 (b) on the main.
}
\end{figure}

\end{document}